\documentclass[]{raa}                  
\usepackage{graphicx,times}             
\usepackage{amsmath}
\usepackage{amsfonts}
\usepackage{amssymb}
\usepackage{multicol}
\usepackage{lineno}
\usepackage{color}
\modulolinenumbers[1]

\begin{document}

\title{Calibration of the DAMPE Plastic Scintillator Detector and its
on-orbit performance \,$^*$ \footnotetext{$*$  Corresponding
Author: y.p.zhang@impcas.ac.cn}}

   \volnopage{Vol.0 (200x) No.0, 000--000}      
   \setcounter{page}{1}          

   \author{Meng Ding
      \inst{1,2}
   \and Yapeng Zhang
      \inst{1}
   \and Yong-Jie Zhang
      \inst{1}
   \and Yuan-Peng Wang
      \inst{3,4}
   \and Tie-Kuang Dong
      \inst{3}
   \and{Antonio De Benedittis}
      \inst{5,6}
   \and Paolo Bernardini
      \inst{5,6}
   \and Fang Fang
      \inst{1}
   \and Yao Li
      \inst{1,2}
   \and Jie Liu
      \inst{1}
   \and Peng-Xiong Ma
      \inst{3,4}
   \and Zhi-Yu Sun
      \inst{1}
   \and Valentina Gallo
      \inst{7}
   \and Stefania Vitillo
      \inst{7}
   \and Zhao-Min Wang
      \inst{8,9}
   \and Yu-Hong Yu
      \inst{1}
   \and Chuan Yue
      \inst{3,4}
   \and Qiang Yuan
      \inst{3}
   \and Yong Zhou
      \inst{1,2}
   \and Yun-Long Zhang
      \inst{10}
   }
   \institute{
             Institute of Modern Physics, Chinese Academy of Sciences, Lanzhou 730000, China \\
        \and
             University of Chinese Academy of Sciences, Yuquan Road 19, Beijing 100049, China \\
        \and
             Key Laboratory of Dark Matter and Space Astronomy, Purple Mountain Observatory, Chinese Academy of Sciences, 2 West Beijing Road, Nanjing 210008, China \\
        \and
             School of Astronomy and Space Science, University of Science and Technology of China, Hefei 230026, China
        \and
             Istituto Nazionale di Fisica Nucleare (INFN) - Sezione di Lecce, I-73100 Lecce, Italy
        \and
             Dipartimento di Matematica e Fisica E. De Giorgi, Universit\`a del Salento, I-73100 Lecce, Italy
        \and
             Department of Nuclear and Particle Physics, University of Geneva, CH-1211 Geneva, Switzerland
        \and
             Gran Sasso Science Institute, Via M. lacobucci 2, 67100 L'Aquila, Italy
        \and
             INFN Laboratori Nazionali del Gran Sasso, 67100 Assergi (L'Aquila), Italy
        \and
             Department of Modern Physics, University of Science and Technology of China, Hefei 230026, China
   }

   \date{Received~~2009 month day; accepted~~2009~~month day}

\abstract{
DArk Matter Particle Explorer (DAMPE) is a space-borne apparatus for detecting the high-energy cosmic-rays like electrons, $\gamma$-rays, protons and heavy-ions. Plastic Scintillator Detector (PSD) is the top-most sub-detector of the DAMPE. The PSD is designed to measure the charge of incident high-energy particles and it also serves as a veto detector for discriminating $\gamma$-rays from charged particles. In this paper, PSD on-orbit calibration procedure is described, which includes five steps of pedestal, dynode correlation, response to minimum-ionizing particles (MIPs), light attenuation function and energy reconstruction. A method for reconstructing the charge of incident high energy cosmic-ray particles is introduced. The detection efficiency of each PSD strip is verified to be above 99.5\%, the total efficiency of the PSD for charged particles is above 99.99\%.
\keywords{Cosmic ray --- Instrumentation: DAMPE --- Charge measurement --- Plastic scintillator detector calibration}
}

   \authorrunning{M. Ding, Y.-P. Zhang  \& Y. Li}            
   \titlerunning{Calibration and charge reconstruction for the DAMPE Plastic Scintillator Detector}  

   \maketitle

%
%
\section{Introduction}           
\label{sect:intro}

\begin{figure}
   \centering
   \includegraphics[width=0.5\textwidth]{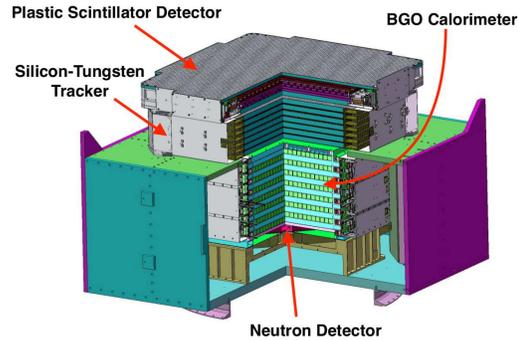}
    \caption{Layout of DAMPE detector}
    \label{Fig:dampe1}
   \end{figure}

Exploring the nature of dark matter has been being one of the most
important topics in fields including cosmology, astrophysics and particle physics. Precisely measuring the energy spectra of cosmic-rays is vital to constrain the cosmic-ray production mechanism (\cite{Luke2012}) and their propagation in the stellar medium (\cite{Isabelle2015}).
Measuring the energy spectrum of cosmic particles like $e^\pm$,
$\gamma$-rays and anti-particles in space is one of the experimental methods to
constrain the properties of dark matter (\cite{Chang2014}).
The space-borne experiment has been pioneered by PAMELA (\cite{pamela}), Fermi-LAT (\cite{fermi}) and AMS-02 (\cite{Alpat2005}).
DArk Matter Particle Explorer (DAMPE) (\cite{changDampe}) is a high-resolution multi-purpose device for detecting cosmic-rays including electrons, $\gamma$-rays, protons and heavy ions in an energy range of a few GeV to 100 TeV. DAMPE has been launched on December 17th, 2015 and operates on a sun-synchronous orbit at the altitude of 500 km.
DAMPE consists of four sub-detectors: a Plastic Scintillator Detector (PSD) (\cite{Yuyuhong}), a Silicon-Tungsten Tracker (STK) (\cite{STK}), a Bismuth Germanate Oxid Calorimeter (BGO) (\cite{zhangzy1, zhangzy2, BGO}) and a NeUtron Detector (NUD) (\cite{NUD}). The structure of DAMPE is shown in Fig.~\ref{Fig:dampe1}.

The on-orbit calibration of PSD is an important step to obtain the charge information of incident particles. In this paper, after a brief introduction (Section {\ref{sec:psdinfo}}) of PSD, PSD on-orbit calibration procedure is described in Section {\ref{sec:psdcali}} including pedestal, dynode, MIPs response, light attenuation and energy reconstruction. A charge reconstruction method and the PSD detection efficiency are presented in Section {\ref{sec:chrg}} and Section {\ref{sec:eff}}, respectively.

\section{Design of the PSD}
\label{sec:psdinfo}
The PSD is designed to fulfill two major tasks:
(a) to measure the charge of incident high-energy particles with the charge number $Z$ from 1 to 26;
(b) to serve as a veto detector for discriminating $\gamma$-rays from charged particles.
All these require the PSD to have a large dynamic range, good energy resolution and high detection efficiency.

\begin{figure}
    \centering
    \includegraphics[width=0.38\textwidth]{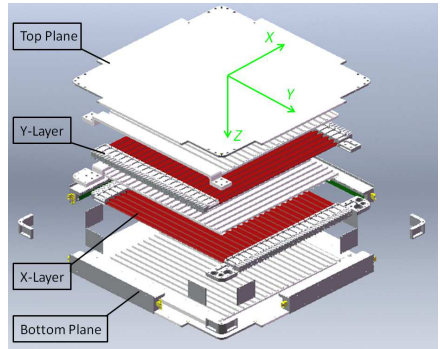}
    \caption{The structure of the PSD.}
    \label{fig:psd}
\end{figure}

\begin{figure}
    \centering
    \includegraphics[width=0.38\textwidth]{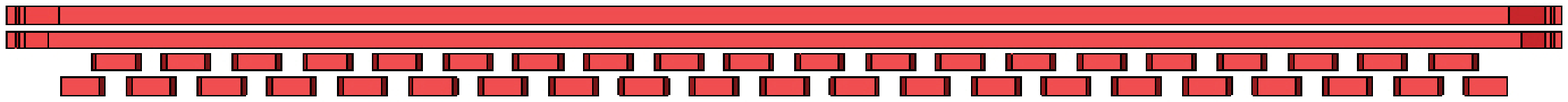}
    \caption{Side view of the PSD bars.}
    \label{fig:psd3}
\end{figure}

The PSD has two layers of plastic scintillator arrays, as shown in Fig.~\ref{fig:psd}. Each layer is composed of 41 plastic scintillator bars and the dimension of bars is 884 mm $\times$ 28 mm (25 mm for bars on edges) $\times$ 10 mm. The bars in top and bottom layers are parallel to the $X$-axis and the $Y$-axis of the DAMPE coordinate system, respectively.
In order to avoid the ineffective detection area, neighboring bars in each layer are staggered by 8 mm, as shown in Fig. \ref{fig:psd3}. The active area of the PSD is 825 mm $\times$ 825 mm.
Scintillation light of each PSD bar is collected by two Photo Multiplier-tubes (PMTs) at two ends. Each PMT (Hamamatsu R334) has 10 dynodes.
In order to cover an energy measurement from 0.1 to 1600 energy deposition of minimum-ionizing protons (EMIPs), signals from dynode 5 and dynode 8 of each PMT are extracted. Signal of dynode 5 with smaller gain covers from 4 EMIPs to 1600 EMIPs and signal of dynode 8 with bigger gain covers from 0.1 EMIPs to 40 EMIPs (\cite{ZhouYong}). Overlapping measurements by two dynodes are designed to calibrate the response of a PMT. More detailed information about the design, assembly and laboratory tests has been described in Refs. (\cite{Yuyuhong}). In order to obtain a stable on-orbit performance of the PSD, an active temperature control strategy was implemented by using front-end electronic boards and additional heat coils as thermal sources (\cite{Yuyuhong}).
The on-orbit temperature variation of the PSD is verified to be less than 1$^o$ C (\cite{LiYao}), which is a crucial factor for maintaining a stable performance of the PSD.

\section{PSD on-orbit calibration}
\label{sec:psdcali}
PSD on-orbit calibration is the procedure to process PSD-associated ADCs to deposited energies with taking noises and non-linearities in actual detection processes into account.
When a charged particle passes through a plastic scintillator bar, the deposited energy would be transformed into scintillation lights. The lights are collected by PMTs attached at both ends, and then the lights are converted into electronic signals. After amplification, shaping and holding etc. processes, the amplitudes of electronic signals are recorded by ADC units.
The PSD calibration procedure is quite similar to reverse these processes, including calibration of pedestal, dynode correlation, response to MIPs, light attenuation function and energy reconstruction. In PSD on-orbit calibration, the data collected in the South Atlantic Anomaly (SAA) region are excluded. Calibration steps mentioned above will be presented in following sub-sections.

\subsection{\label{sec:pedestal1} Pedestal calibration}
\begin{figure}
    \vspace{-2mm}
    \centering
    \includegraphics[width=7cm,height=5cm]{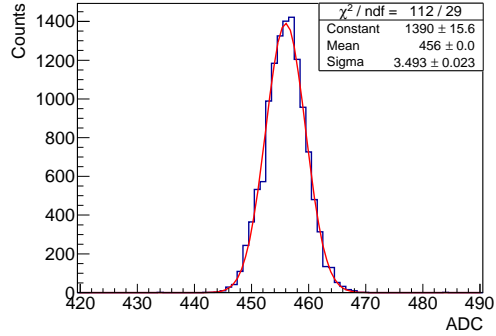}
    \caption{A typical pedestal distribution of a readout c/hannel, the fitted Gaussian function is depicted by the red solid curve.}
     \label{fig:ped}
\end{figure}

Pedestal of each readout channel is sampled twice per orbit under random triggers.
The pedestal distribution of each channel is a Gaussian-like distribution. Fig.~\ref{fig:ped} shows a typical pedestal distribution of a readout channel, the Gaussian fit-function is depicted by the red solid curve.
Obtained pedestals (mean value of the fitted Gaussian function) and
standard deviation (Gaussian $\sigma$) of each readout channel of the
PSD are written into a pedestal calibration file with a time tag associated to
the input data.

\begin{figure}
    \centering
    \includegraphics[width=0.60\textwidth]{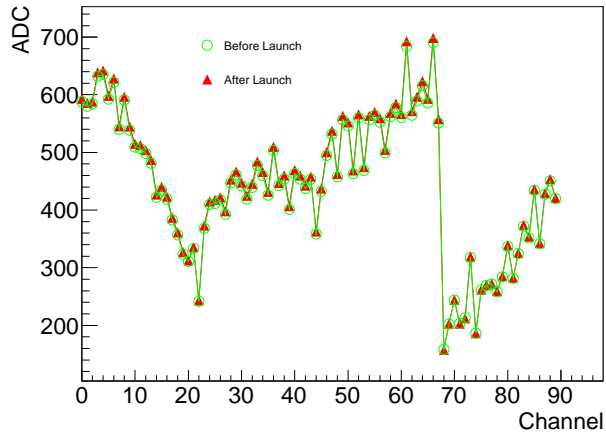}
    \caption{Pedestal before (green circles) and after (red triangles) the launch of DAMPE.}
    \label{fig:pedcomp}
\end{figure}

\begin{figure}
    \centering
    \includegraphics[width=0.60\textwidth]{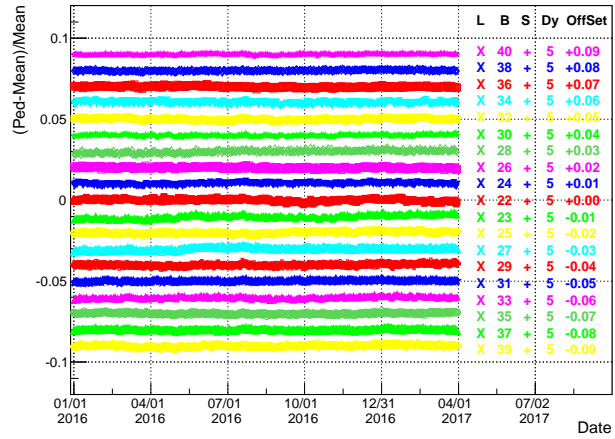}
    \caption{Percentage variation of pedestals versus date (an offset is introduced for display reason).}
    \label{fig:pedDate}
\end{figure}

PSD on-orbit pedestals are almost the same as those measured before the launch of DAMPE. As an example, Fig.~\ref{fig:pedcomp} shows pedestals of all readout channels of the positive side of X-layer (including 4 spare channels) before (circles) and after (triangles)
the launch of the DAMPE.
In order to characterize the on-orbit pedestal stability of the PSD, a percentage variation of pedestal is defined as ($PED$-Mean)/Mean, where $PED$ is the pedestal of a readout channel on different date and Mean is the average pedestal of the same channel. Fig. ~\ref{fig:pedDate} shows the percentage variation of pedestals versus date, each horizontal line represents a readout channel. The corresponding information of the channel, i.e. layer (L), bar (B), side (S), dynode number (Dy) and offset are listed at the right side of the figure. The offset is introduced for display reason only.
In general, the pedestals of all PSD readout channels are stable and the overall variation is less than 0.1\% during the first 18 months of data-taking. PSD pedestals are daily updated, and then they are subtracted from the on-orbit data.

\subsection{\label{sec:dy58ratios} Dynode calibration}
\begin{figure}
    \centering
    \includegraphics[width=0.40\textwidth]{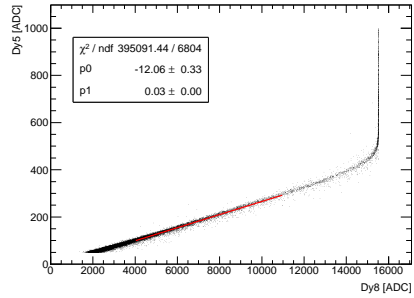}
    \caption{Dy5 (vertical axis) and Dy8 (horizontal axis) of a  PSD PMT, the red line is the linear fit-function in the first iteration.}
    \label{fig:Dy58_1}
\end{figure}

\begin{figure}
    \centering
    \includegraphics[width=0.40\textwidth]{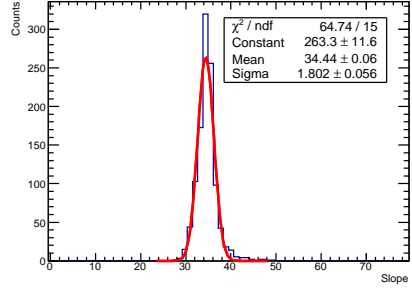}
    \caption{Distribution of calculated slope parameters ($k_1$) of a PMT in the second iteration.}
    \label{fig:Slp}
\end{figure}

As mentioned in section {\ref{sec:psdinfo}}, each PSD PMT gives two signals by
its dynode 5 (Dy5) and dynode 8 (Dy8) in order to achieve a large dynamic range.
Typically, ADC values of Dy5 and Dy8 of a PMT have a linear correlation before the
Dy8 gets saturated.
In order to obtain the correlation of Dy5 and Dy8 with less bias, the dynode calibration is made via a two-step iterations. In the first iteration, the correlation of Dy8 and Dy5 for each PMT is built, and it is fitted with a linear function. A typical correlation of Dy8 and Dy5 of a PMT is shown in Fig.~\ref{fig:Dy58_1}, the linear fit-function is depicted by the red line, where p0 and p1 are fitting parameters.
Then, Dy8 is expressed as a linear function of Dy5, the slope parameter ($k_0$=1/p0) and the intercept parameter ($b_0$=-p1/p0) are obtained (the reason for obtaining values of above parameters in such way is that the fitting range for Dy8 is more straightforward than the one of Dy5).
In second iteration, the slope parameter of the dynode correlation is calculated by $k_1$=(Dy8-$b_0$)/Dy5. Fig.~\ref{fig:Slp} shows the distribution of calculated slope parameters ($k_1$) of a PMT, which is fitted with a Gaussian distribution shown
by the red solid curve. The mean ($k_{mean}$) and the standard deviation
($\sigma_k$) of the fitted Gaussian function are extracted.
In order to exclude abnormal events, the events within $|k_1-k_{mean}|<5\sigma_k$ are selected, by using which the slope parameter $k$ and intercept parameter $b$ are obtained for each PMT just liking the first iteration.

%
With the final obtained dynode calibration parameters, two ADC measurements of each PSD PMT are combined into a single ADC value, which is expressed as Eq.{~\ref{eq:dy58}},
\begin{equation}
    \text{ADC}=
    \begin{cases}
         ADC_{\text{Dy8}}          & ADC_{\text{Dy8}} \leq 11000 \\
         k{\times}ADC_{\text{Dy5}}+b & ADC_{\text{Dy8}} > 11000
    \end{cases}
    \label{eq:dy58}
\end{equation}
where $ADC_{\text{Dy5}}$ and $ADC_{\text{Dy8}}$ are ADC values associated to Dy5 and Dy8 of a PMT, respectively.

\begin{figure}[h]
  \begin{minipage}[t]{0.495\linewidth}
  \centering
  \includegraphics[width=59mm,height=50mm]{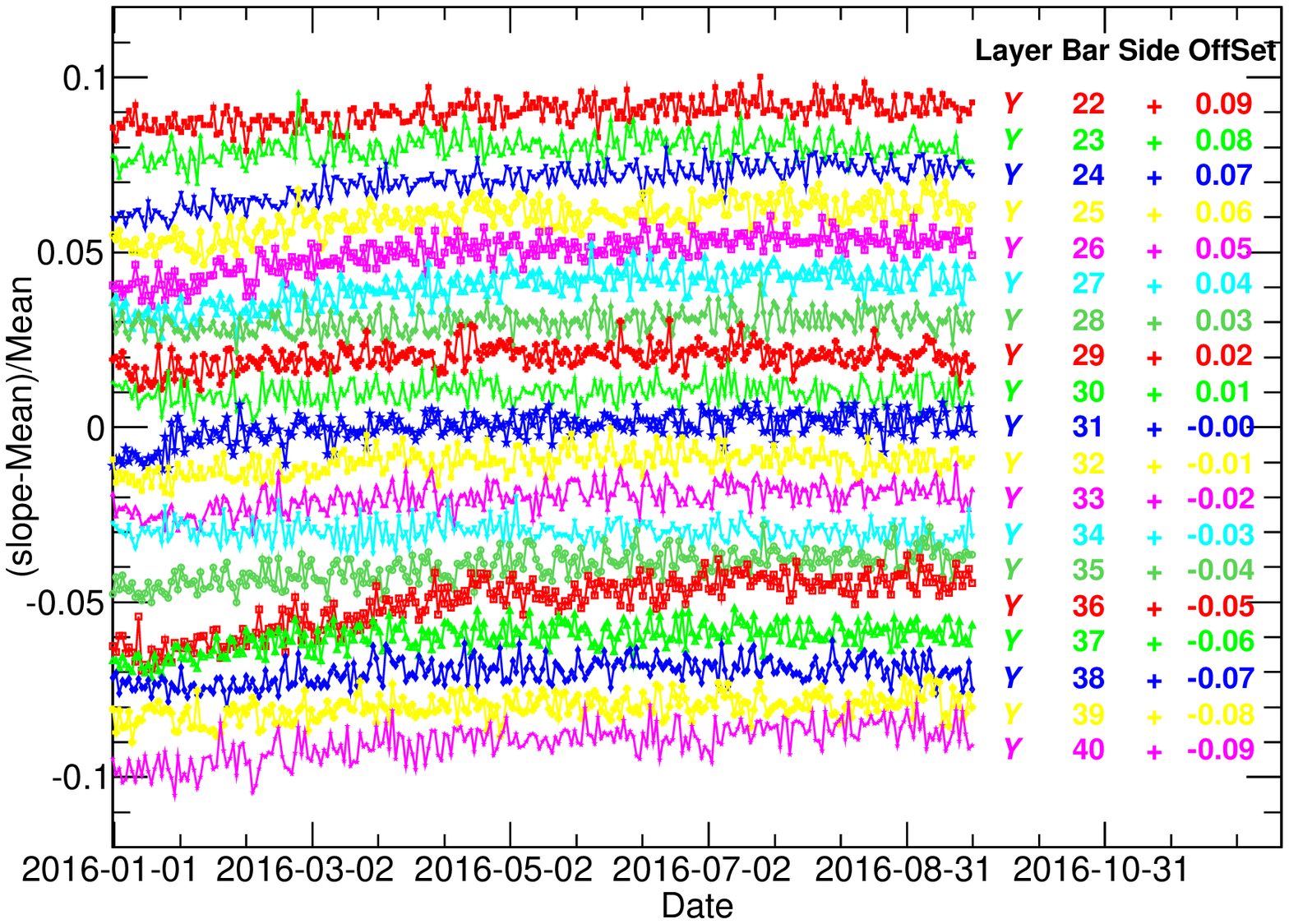}
  \end{minipage}%
  \begin{minipage}[t]{0.495\textwidth}
  \centering
   \includegraphics[width=59mm,height=50mm]{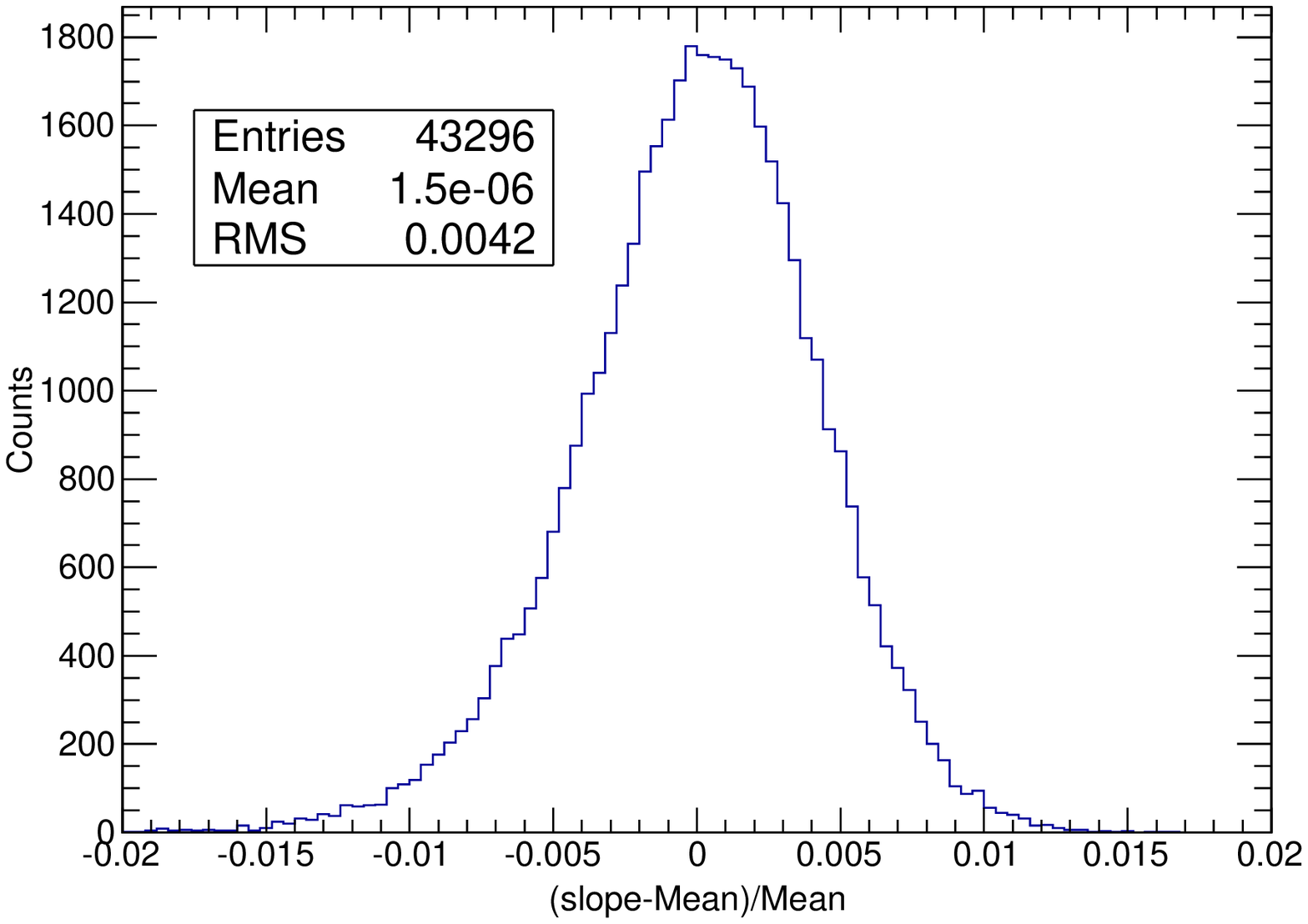}
  \end{minipage}%
  \caption{Left panel: Percentage variation of the slope parameter of channels versus the
time. Right panel: the overall distribution of the percentage variation of the slope
parameter.}
   \label{fig:Dy58}
\end{figure}

The left panel of Fig.~\ref{fig:Dy58} shows the percentage variation of the slope parameter versus time. The percentage variation of the slope parameter is defined just like the one of the pedestal (see Fig. \ref{fig:pedDate}). The slope percentage variation of all channels listed in the left panel of Fig.~\ref{fig:Dy58} is shown in the right panel of Fig.~\ref{fig:Dy58}, the overall change of the slope parameter in this time period is about 0.42\%.
The dynode calibration is made on a daily base. In this step, the linearity of Dy8 can be controlled by the measurements from Dy5, while the Dy5 may also get non-linear at large ADC values, when Dy8 is saturated. The influence of this effect will be further considered in Section~{\ref{sec:chrg}}.

%
\subsection{Response of MIPs and energy reconstruction}
\label{sec:mips}

Each PSD bar provides two ADC measurements by its left and the right side ($ADC^{L/R})$ after the dynode calibration, a combined ADC of each PSD bar is constructed via the geometrical average of $ADC^L$ and $ADC^R$, i.e. $ADC^C=\sqrt{ADC^L{\times}ADC^R}$.
Due to the light attenuation in the scintillator bar, the combined quantity $ADC^C$ has less hit position dependence than the singe-side quantities. This effect will be illustrated in following sub-sections.

\begin{figure}
    \centering
    \includegraphics[width=0.50\textwidth]{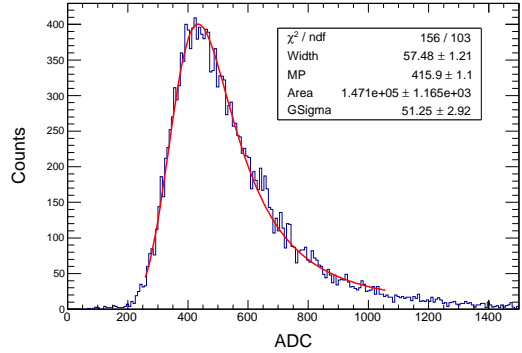}
    \caption{Typical ADC distribution of MIP, the red curve is the fitted Landau distribution convoluted with a Gaussian function.}
    \label{fig:mipsadc}
\end{figure}

\begin{figure}
    \centering
    \includegraphics[width=0.50\textwidth]{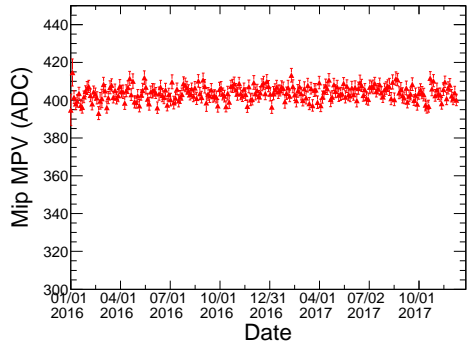}
    \caption{Fitted MPV value of a PMT versus the time.}
    \label{fig:MipDate}
\end{figure}

The events are classified as MIP events, if incident particles pass through all BGO layers and their energy depositions in each BGO layer are in the range expected for MIPs (\cite{DmpMIPs}).
Fig.~\ref{fig:mipsadc} shows a typical ADC distribution of MIP events with path length correction (PSD alignment method will be published in another paper). Considering the detector resolution, the distribution is fitted by a Landau distribution convoluted with a Gaussian function (LG), shown by the red solid curve.
The most probable value (MPV) of the fitted Landau distribution is obtained.
For each PMT, MIPs responses in each PSD bar, i.e. MPV$^L$, MPV$^R$ and MPV$^C$, are
obtained by fitting the corresponding $ADC^{L/R/C}$ distribution with the LG function.
The MPV values of each PSD bar are updated every five days.
Thanks to the temperature control system, MPV values of PSD PMTs are very stable. Fig. ~\ref{fig:MipDate} shows MPV of a PMT versus the time, MPV variation of this PMT is less than 4 ADC channels.

Considering the energy deposition of minimum-ionizing protons passing through the
plastic scintillator is about 2 MeV/cm (the thickness of a PSD bar is 1 cm).
The measured energies of left/right/combined side of each PSD bar
($E^{L/R/C}$) are derived according to
\begin{equation}
E_{i}^{L/R/C} =  \frac{ADC^{L/R/C}_i}{MPV^{L/R/C}_i} \times 2~\text{MeV},
\label{eq:Erec}
\end{equation}
where, $ADC^{L/R/C}_{i}$ are ADC value of left/right/combined side of
$i$-th PSD bar.

\subsection{Light attenuation calibration}
\label{sec:Att}
The scintillation lights generated along the path of incident charged particle would attenuate during their propagation due to light absorption and re-scattering. In general, the amount of scintillation lights received by a PMT is inversely proportional to the distance between hit position and the PMT.
Due to the light attenuation effect, the obtained energy (Eq. \ref{eq:Erec}) needs to be corrected according to corresponding light attenuation function and the hit position.

\begin{figure}[h]
    \begin{minipage}[t]{0.5\linewidth}
    \centering
    \includegraphics[width=59mm,height=50mm]{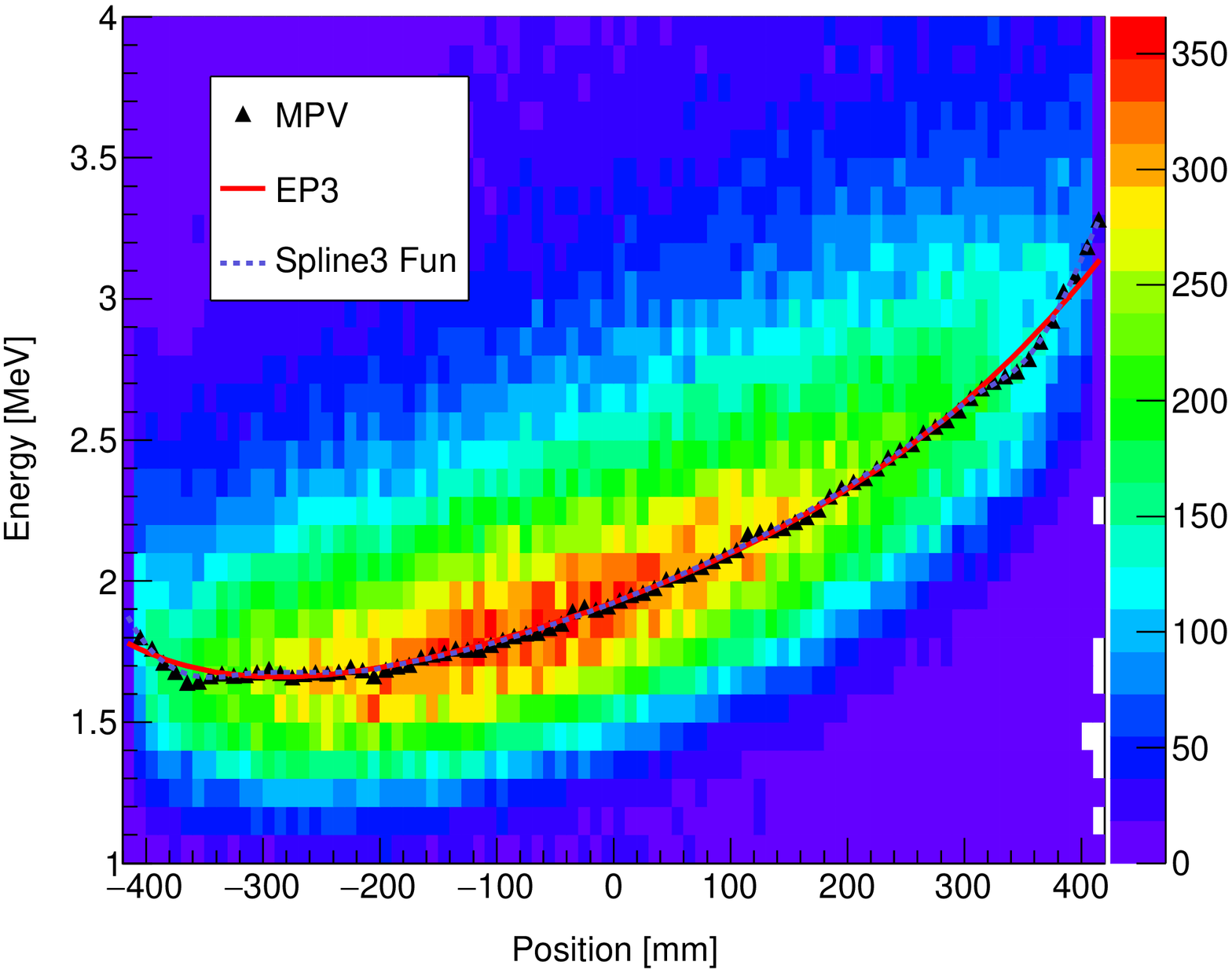}
    \caption{Energy of minimum-ionizing protons (vertical axis) versus the hit position (horizontal axis). The triangles are the MPV values of the energy distributions for the bins of the horizontal axis, the red solid line is the fit-function as Eq.~\ref{eq:psdatt} and the blue dashed line is the constructed 3rd-spline function.}
    \label{fig:attcomp}
    \end{minipage}%
    \begin{minipage}[t]{0.5\textwidth}
    \centering
    \includegraphics[width=59mm,height=50mm]{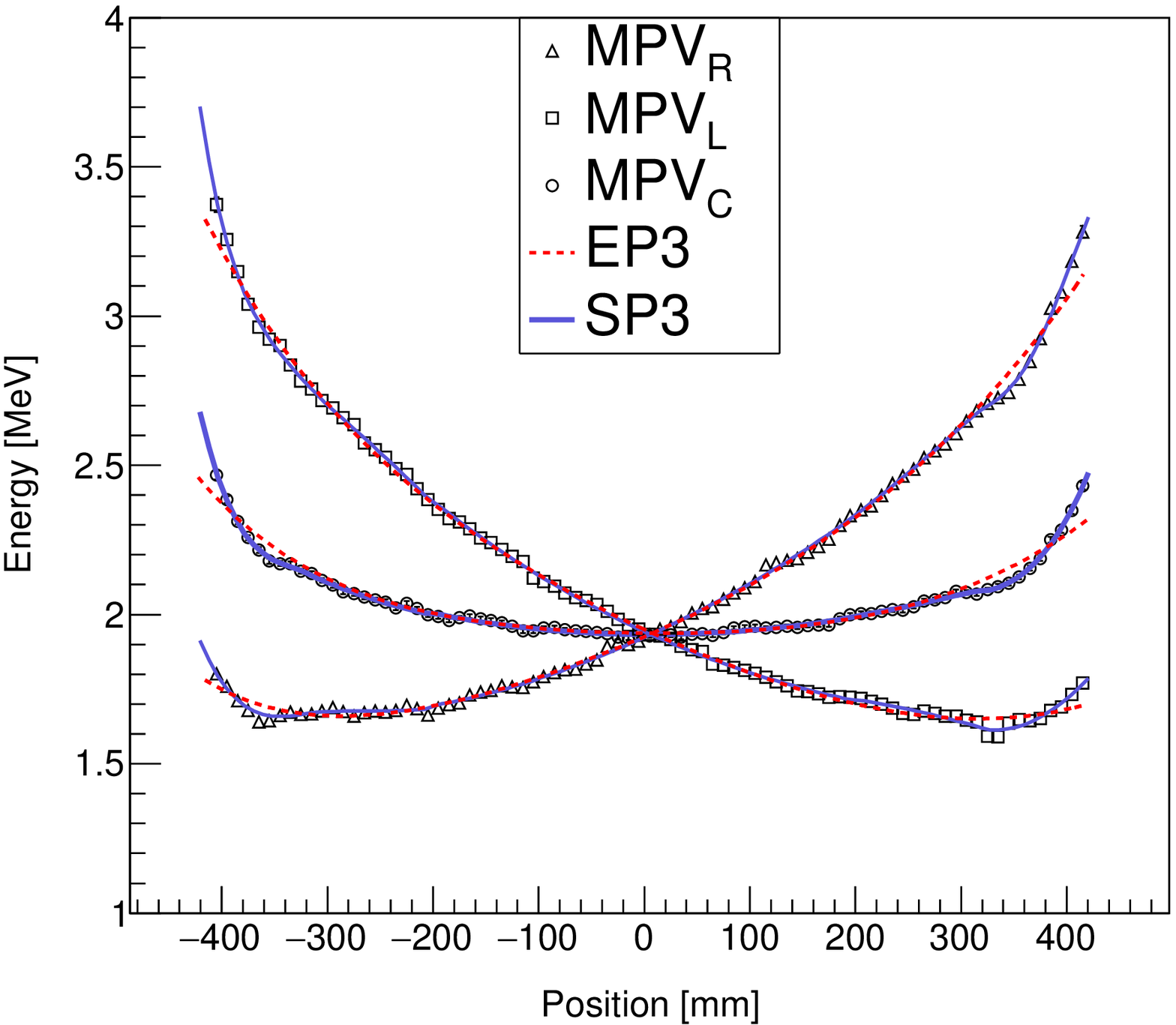}
    \caption{MPVs (vertical axis) of right (triangles), left (circles) and combined (squares) side of a PMT versus hit position (horizontal axis). The fitted functions as Eq.~\ref{eq:psdatt} and
constructed spline functions are shown by the red dashed lines and the blue
solid lines, respectively.}
     \label{fig:att3}
    \end{minipage}%
\end{figure}

The light attenuation behavior of each PSD bar is investigated using a sample of MIP events with an unambiguous global track (global track defined as an STK track compatible with BGO track), the hit positions are obtained by extrapolating the track to PSD sub-layers. Fig.~\ref{fig:attcomp} shows a typical scatter plot of reconstructed energy of a PMT with path length correction
and obtained hit position. Energy distributions of each horizontal bin of Fig.~\ref{fig:attcomp} are obtained, and they are fitted by a Landau distribution, respectively. The obtained MPV values
are shown by triangles in Fig.~\ref{fig:attcomp}.
The light attenuation functions of left/right side of PSD strips are
obtained by fitting the corresponding correlation of MPV versus hit position with a function in Eq. ~\ref{eq:psdatt}, which is the linear combination of an exponential function and a 3rd-order polynomial function (EP3),
\begin{equation}
\label{eq:psdatt}
     A(x) = C_0e^{-x/\lambda}+C_1+C_2x+C_3x^2+C_4x^3~,
\end{equation}
where $\lambda$ and $C_0,...C_4$ are fitting parameters and $x$ is the
hit position. The fit function is depicted by the red curve in Fig.~\ref{fig:attcomp}.

In order to take into account possible large structure(s) in the correlation of MPV versus hit position, the correlations of left or right side of PSD bars are smoothed by fourth-order polynomial functions in four regions (range of neighboring regions are overlapped), respectively. The smoothed MPV at each hit position is obtained from one of smoothing polynomial functions.
Based on smoothed MPV data and hit positions, attenuation functions of each side of PSD bars are constructed by means of the 3rd-order spline method (ROOT::TSpline). Light attenuation function for combined energy of a PSD bar (see Eq.~\ref{eq:Erec}) is calculated by attenuation functions of left side and right side, i.e. $A^C(x)=\sqrt{A^L(x){\times}A^R(x)}$.
As an example, typical light attenuation functions (MPV value versus hit position) of left (triangles), right (squares) and combined (circles) sides of a PSD bar are shown in
Fig.~\ref{fig:att3}, the fitted EP3 functions and spline functions (noted as SP3) are shown by the dashed and solid line, respectively. Both Eq.~\ref{eq:psdatt} and Spline
function can well describe the light attenuation behaviors of PSD bars.

\section{Charge reconstruction}
\label{sec:chrg}
According to the Bethe equation (\cite{PDG2016}), the energy loss of
charged particles in the matter is proportional to the square of their electric
charge, and it is increasing slowly with the energy when $\beta\gamma>4$. Therefore, the charge of incident particle can be obtained by comparing its energy deposition to the one of minimum-ionizing protons.

The reconstructed charge of incident particles ($Q_{rec}^{L/R/C}$) could be extracted by following expression:
\begin{equation}\label{eq:QRec}
     Q_{rec}^{L/R/C} =
\sqrt{\frac{E^{L/R/C}}{A^{L/R/C}(x)}\times\frac{\text{S}}{l}}~,
\end{equation}
where, $E^{L/R/C}$ is the energy of left/right/combined
side of a PSD bar, $l$ is the path length of the particle inside the volume of the PSD bar, S=10 mm is the thickness of the PSD bar, $A^{L/R/C}(x)$ is the corresponding light attenuation function (see Eq.~\ref{eq:psdatt} and Fig. \ref{fig:attcomp}) and $x$ is the hit position given by the track.

\begin{figure}
    \centering
    \includegraphics[width=0.5\textwidth]{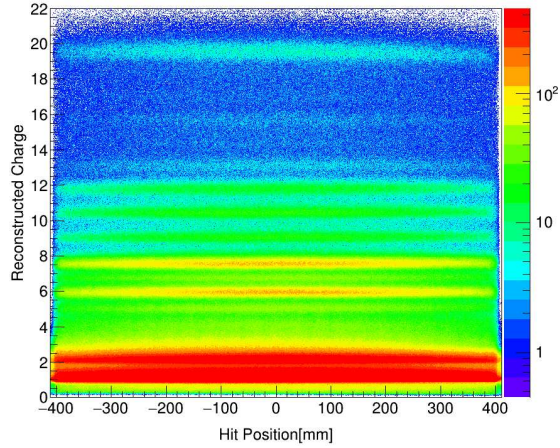}
    \caption{Reconstructed charge (vertical axis) versus the hit position (horizontal axis).}
    \label{fig:qpos}
\end{figure}

Fig.~\ref{fig:qpos} shows a correlation of the reconstructed combined-side charge ($Q^C_{rec}$) of the X-layer of the PSD and the hit position. The horizontal bands represent cosmic ray nuclei with different charges.
In Fig.~{\ref{fig:qpos}}, reconstructed charges (different bands) at large values are not at their nominal charges. For example, the top band is corresponding to Fe (Z=26) while the band locates at about 20. This is mainly due to the so-called quenching effect (\cite{Briks1964}), and partially due to the gain non-linearity of individual channel. The quenching effect represents the fact that the energy deposition of high Z particles in matter is smaller than the one given by the Bethe equation.
Besides, signals from high Z particles are measured by the Dy5 of PSD PMTs, the gain linearities are slightly different from one readout channel to another. Due to these reasons, the charge spectra reconstructed by different PMTs may vary slightly.

\begin{figure}
    \centering
    \includegraphics[width=9cm]{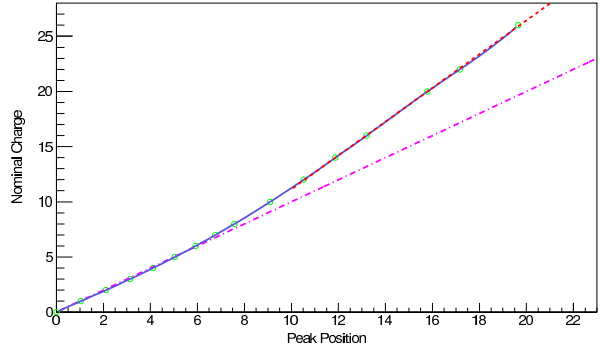}
    \caption{Peak positions obtained from a reconstructed charge spectrum versus charge numbers (open circles). The solid line and dotted line present the constructed third-order spline function and the fitted linear function, respectively, the dash-dotted line is the bisector (ordinate = abscissa) as a reference.}
    \label{fig:Que}
\end{figure}

In order to correct the abovementioned effects, visible peaks in reconstructed charge spectra ($Q^{L/R/C}$ distribution) of each PSD bar are fitted by a series of Gaussian fits, and peak positions (mean of Gaussian function) are obtained, respectively. The peaks are assigned to known cosmic-ray nuclei based on the knowledge of cosmic nuclei abundance.
Based on obtained peak position and nominal charge number pairs, a third-order spline function (SP3) for each charge measurement (left/right/combined side) of a PSD bar is constructed, respectively. 
For each reconstructed charge (see Eq. \ref{eq:QRec}), a quenching-effect corrected charge can be obtained from the corresponding third-order spline function. For reconstructed charge above Fe i.e. $Q_{rec}>Q^{\text{Fe}}_{rec}$, a linear function is used to correct the quenching effect, which is fitted from the pairs of peak position and nominal charge number with $Q_{rec}>10$. After applying this correction, the charge spectra reconstructed from different PSD bars are aligned on a bisector ($Q_{rec}$=Z).
As an example, Fig. ~\ref{fig:Que} shows a typical data pairs of obtained peak position and charge number (open circles),  the constructed third-order spline function and the fitted linear function are depicted by the solid line and dotted line, respectively. The dash-dotted line is the bisector (ordinate = abscissa) as a reference.

As shown by the Eq.~\ref{eq:QRec}, the hit position ($x$) and the path length ($l$) are derived from a selected track. It is common that the track number of an event is more than one, the situation is much more complicated for the case of high-Z nuclei events. In order to select a right track, one needs to combine the information provided by PSD, STK and BGO together. Detailed track selection method and reconstruction results will be published in another paper.

\section{PSD detection efficiency}
\label{sec:eff}
Electromagnetic showers can be well separated from the hadronic showers by the BGO calorimeter (\cite{DmpNature2017}). The flux of the cosmic $\gamma$-ray is about 1000 times lower than the one of the electrons. Therefore, the detection efficiency of the PSD is crucial for discriminating the $\gamma$-ray from charged particles.
By combining measurements from different sub-detectors, the detection efficiency
of each PSD strip can be evaluated. The PSD detection efficiency is investigated
by MIP events as well.
The detection efficiency of a PSD bar $\eta_{l,b}$ ($l$ and $b$ represent the layer
number and strip number, respectively) is defined as following,
\begin{equation}\label{f:psdeff}
\eta_{l,b} = N^{Fired}_{l,b}/N^{STK}_{l,b}~,
\end{equation}
where $N^{STK}_{l,b}$ is the number of the global track pointing to $b$-th
strip
in $l$-th layer and $N^{Fired}_{l,b}$ is the number of the fired PSD bar.
The firing condition of a PSD bar is that energies with light attenuation correction in left and right side of the bar are larger than 0.2 MeV.

\begin{figure}
    \centering
   \includegraphics[width=0.50\textwidth]{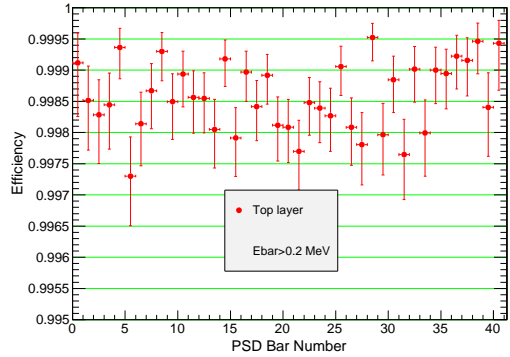}
    \caption{The detection efficiency of bars of PSD top layer.}
    \label{fig:psdeff}
\end{figure}

In PSD efficiency evaluation, MIPs events which fulfill following conditions
are used:
a) Event with only one global track; b) the STK track has 5 clusters in both the $X-Z$
plane and the $Y-Z$ plane at least; c) The reduced $\chi^2 (\chi^2/NDF)$ is smaller than
1; d) STK track and BGO track are in good agreement.
Fig.~\ref{fig:psdeff} shows the detection efficiency of bars in the top
layer of the PSD.
The efficiencies of all bars in the X-layer and the Y-layer of the PSD are readily above 99.5\%, which is far better than the designed detection efficiency of 95\%.
The total efficiency of the PSD (at least one PSD bar is fired in top layer or
 bottom layer) is evaluated to be above 99.99\%.

\section{\label{sec:summary} Summary}

PSD on-orbit calibration procedure including five steps of pedestal, dynode correlation, MIP response, light attenuation and energy reconstruction are presented in this paper.
The on-orbit pedestal, dynode ratio and MIPs response of the PSD are verified to be stable.
A method for reconstructing the charge of incident particle is introduced including quenching effect and gain non-linearity corrections.
The detection efficiency of the PSD is evaluated with MIPs, the detection efficiency of each PSD bar is large than 99.5\%, the total detection efficiency of the PSD is above 99.99\%.

\begin{acknowledgements}
This work was funded by National Key ggProgram for Research and Development (2016YFA0400201), also supported by National Natural Science Foundation of China (11673047, 11673075, 11303107, U1738127 and U1738205).
\end{acknowledgements}

\label{lastpage}

\end{document}